\documentclass[sigconf]{acmart}
\usepackage{multirow}
\usepackage{svg}
\usepackage{subcaption}
\usepackage{threeparttable}
\usepackage[normalem]{ulem}
\usepackage{enumitem}
\useunder{\uline}{\ul}{}
\newtheorem{mydef}{Definition}
\AtBeginDocument{%
  \providecommand\BibTeX{{%
    \normalfont B\kern-0.5em{\scshape i\kern-0.25em b}\kern-0.8em\TeX}}}
\setcopyright{acmcopyright}
\copyrightyear{2024} 
\acmYear{2024} 
\setcopyright{acmlicensed}\acmConference[WSDM '24]{Proceedings of the Seventeenth ACM International Conference on Web Search and Data Mining}{March 4-8, 2024}{Merida, Mexico}
\acmBooktitle{Proceedings of the Seventeenth ACM International Conference on Web Search and Data Mining (WSDM '24), March 4-8, 2024, Merida, Mexico}
\acmPrice{15.00}
\acmDOI{10.1145/xxxxxxx.xxxxxxx}
\acmISBN{978-1-4503-9407-x/xx/xx}

\begin{document}

\title{Unified Pretraining for Recommendation via Task Hypergraphs}

\author{Mingdai~Yang}
\affiliation{%
  \institution{University of Illinois at Chicago}
  \city{Chicago}
  \country{USA}}
\email{myang72@uic.edu}

\author{Zhiwei~Liu}
\affiliation{%
  \institution{Salesforce AI Research}
  \city{Palo Alto}
  \country{USA}
}
\email{zhiweiliu@salesforce.com}


\author{Liangwei~Yang}
\email{lyang84@uic.edu}
\affiliation{%
  \institution{University of Illinois at Chicago}
  \city{Chicago}
  \country{USA}}

\author{Xiaolong~Liu}
\email{xliu262@uic.edu}
\author{Chen~Wang}
\email{cwang266@uic.edu}
\affiliation{%
  \institution{University of Illinois at Chicago}
  \city{Chicago}
  \country{USA}}

\author{Hao Peng}
\affiliation{%
   \institution{School of Cyber Science and Technology, Beihang University,}
   \country{Beijing, China}\\
   \institution{Yunnan Key Laboratory of Artificial Intelligence, Kunming University of Science and Technology,}
   \country{Kunming, China}}
\email{penghao@buaa.edu.cn}
\authornote{Corresponding author}

\author{Philip S.~Yu}
\affiliation{%
  \institution{University of Illinois at Chicago}
  \city{Chicago}
  \country{USA}}
\email{psyu@uic.edu}


\begin{abstract}
Although pretraining has garnered significant attention and popularity in recent years, its application in graph-based recommender systems is relatively limited. It is challenging to exploit prior knowledge by pretraining in widely used ID-dependent datasets. On one hand, user-item interaction history in one dataset can hardly be transferred to other
datasets through pretraining, where IDs are different. On the other hand, pretraining and finetuning on the same dataset leads to a high risk of overfitting.
In this paper, we propose a novel multitask pretraining framework named \textbf{U}nified \textbf{P}retraining for \textbf{R}ecommendation via \textbf{T}ask \textbf{H}ypergraphs. For a unified learning pattern to handle diverse requirements and nuances of various pretext tasks, we design task hypergraphs to generalize pretext tasks to hyperedge prediction.
A novel transitional attention layer is devised to discriminatively learn the relevance between each pretext task and recommendation.
Experimental results on three benchmark datasets verify the superiority of UPRTH. Additional detailed investigations are conducted to demonstrate the effectiveness of the proposed framework.

\end{abstract}

\ccsdesc[500]{Information systems~Collaborative filtering}
\keywords{Recommender System; Multitask Pretraining; Hypergraph Learning;}


\maketitle

\section{Introduction}

Multitask pretraining has gained significant attention in the field of recommender systems. On the one hand, since users and items are usually tied with unique IDs in each dataset for recommendation, the prior knowledge obtained by pretraining on only user-item interactions in a dataset can hardly be transferred to other downstream datasets where IDs are different. On the other hand, if pretraining and finetuning are performed on user-item interactions in the same dataset, the model has a high risk of memorizing the specific patterns and distributions of the pretraining data and being affected by the overfitting problem.
On the contrary, when a model is pretrained on multiple related tasks simultaneously, it is exposed to diverse contextual cues for personalized recommendation. As a result, the model is encouraged to discover relevant commonalities and underlying structures across various tasks, leading to improved generalization capabilities and robust representations.

Previous works applying multitask pretraining for recommendation rely on language models to capture contextual information from text. P5~\cite{p5} converts all data, such as user-item interactions, item metadata, and user reviews, to natural language sequences and then pretrains on a Transformer~\cite{transformer} model that produces target responses according to textual inputs. Analogously, M6-Rec~\cite{m6rec} converts all tasks in recommender systems into language understanding and language generation, and leverages a generative pretrained language model M6~\cite{m6} to predict user preferences as plain texts. These pretraining frameworks based on language models can unify various subtasks and exploit abundant semantics inside the training corpora for recommendation.

However, obvious linguistic biases are observed in the pretraining frameworks based on language models. In other words, such frameworks prefer predicting fluent and grammatically correct text to recommending items based on user preferences~\cite{Zhang2021}. In addition, unifying subtasks by converting all data to plain text inevitably introduces extra noise into original structure information, which is crucial in capturing user preference through collaborative filtering~\cite{glrs_survey}. Besides these drawbacks, the lack of raw plain text data in datasets further prevents the wide application of language-based multitask pretraining frameworks, although language models support models with sufficient flexibility to unify diverse pretext tasks during pretraining.

Even so, designing a unified learning pattern that effectively handles the diverse requirements and nuances of each task is still important for a pretraining framework. Unifying the learning across various tasks allows the framework to develop shared representations capturing common patterns and features across the tasks. Tasks with similar underlying structures or require reasoning abilities can benefit from shared representations capturing common knowledge. Moreover, unification across tasks helps address the issue of data sparsity for individual tasks. When certain tasks have limited labeled data, the shared knowledge learned from other tasks can be transferred to improve performance.

To both preserve the original structure information in pretext tasks and realize the unification across tasks, we propose a novel framework called Unified Pretraining for Recommendation via Task Hypergraphs (UPRTH). Above all, the hypergraph is an irreplaceable data structure to build a unified architecture for multitask pretraining.
Although edges in a graph are widely used to represent pairwise user-item interactions in recommender systems, they can hardly express more complex high-order relations in other tasks, such as multiple items in the same category for category classification tasks, or multiple users in the same group for group identification tasks. On the contrary, the hyperedge simultaneously connecting multiple nodes is more flexible to represent various relations in different pretext tasks.
In order to ensure the unification across pretext tasks in pretraining, we generalize all pretext tasks to the prediction of different hyperedges in task hypergraphs. For example, for an item category classification task as a pretext task, we regard each item as a node and each category as one hyperedge, and then pretrain the model to predict if an item in a specific category is connected to the corresponding hyperedge. 

Next, to learn prior knowledge from the original structure information during the pretraining on task hypergraphs, only the embeddings of user and item nodes are learned to integrate all the contextual information from pretext tasks into user and item representations. In this way, user preferences and item properties reflected by different pretext tasks are completely preserved in these embeddings for the following finetuning. 

Within a unified pretraining framework, a formidable challenge lies in balancing the degrees of relevance between auxiliary tasks and the target recommendation task during pretraining. Certain tasks may be more closely related to the recommendation problem, while others might be less informative or even potentially detrimental. As a result, it becomes essential to determine the appropriate level of knowledge transfer from each task to ensure that the most relevant and useful information is incorporated for personalized recommendations. Additionally, while the framework should learn generalized representations from multitask pretraining to enhance generalization, it must also capture task-specific nuances to provide accurate and personalized recommendations. Striking the right balance between these two aspects is crucial to avoid overgeneralization on the downstream task.

To adaptively learn the degrees of relevance between each auxiliary task and the target recommendation task during pretraining, we propose a novel transitional attention layer (TA). This attention mechanism ensures the discriminative incorporation of the information from various pretext tasks for the target recommendation task.
Additionally, this TA layer directly transmits task-specific knowledge of users and items in auxiliary tasks to the recommendation task during multitask pretraining, which encourages the framework to capture task-specific nuances to avoid overgeneralization. 

The main contributions of this paper are summarized as follows:
\begin{itemize}[leftmargin=*]
    \item  We design a novel framework UPRTH, a model-agnostic unified pretraining framework based on hypergraph learning for recommendation. To the best of our knowledge, this is the first attempt toward unified multitask pretraining for recommendation, achieved by generalizing all pretext tasks to task hypergraphs.
    \item We devise a novel TA layer to harness the knowledge from each pretext tasks, which discriminatively learns the degrees of relevance between each auxiliary task and the target recommendation task during pretraining.
    \item We conduct extensive experiments on three real-world datasets. The significant improvement of UPRTH on all datasets indicates its superiority as a pretraining framework for recommendation.
\end{itemize}

\section{Related works}
\subsection{Hypergraph Learning for Recommendation}
The application of hypergraph learning has proven successful in various graph-based tasks~\cite{feng19,gao22}. Consequently, researchers have also explored the application of hypergraphs in recommender systems, recognizing their ability to capture complex high-order dependencies through hyperedge-node connections~\cite{hgcncc,hccf,dhcf,gtgs}.

Thanks to the flexibility of hyperedge construction, some previous works leverage hypergraph learning to integrate contextual information in addition to user-item interaction history for personalized recommendation. MRH~\cite{mrh} constructs hyperedges among users, music, and music tags to create a hypergraph for music recommendation. MHCN~\cite{mhcn} introduces a multi-channel hypergraph convolutional network that leverages high-order user-user relations to enhance social recommendation. 
To the best of our knowledge, no previous method has applied hypergraph to unify various pretext tasks during pretraining.

\subsection{Graph Neural Networks with Pretraining}
Graph Neural Networks (GNNs) have garnered significant attention due to the prevalence of graph-structured data in various domains. In order to enhance learning effectiveness on graphs, researchers have delved into the exploration of pretraining GNNs using unlabeled graph data. 
GCC~\cite{gcc} applies contrastive learning to capture the universal topological properties across multiple networks. 
GPPT~\cite{gppt} leverages a graph prompting function to enhance the reliability and efficiency of pretrained GNNs. 
GPT-GNN~\cite{gptgnn} introduces an attributed graph generation task in pretraining to capture the semantic properties of the graph.

Most existing pretrained GNN frameworks are designed and optimized for classification tasks as downstream tasks~\cite{Jiang21,l2pgnn, gcc}. These classification frameworks heavily depending on node attributes during the finetuning process are not well-suited for recommendation when attributes are unavailable in the downstream recommendation task.

\subsection{Recommender Systems with Pretraining}
Recommender systems play a crucial role in helping users discover relevant items from vast collections~\cite{lightgcn20,cfag,ziwei23}. Recent studies have explored the integration of pretraining techniques to enhance the performance of recommender systems~\cite{Xu2022,yuwei23}. One notable approach in this context is self-supervised pretraining with data augmentation
~\cite{sgl,s3rec}, 
which offers a promising avenue for leveraging large amounts of unlabeled data as the source of prior knowledge.

Given the increasing abundance of contextual data in recommender systems, researchers have begun to explore other pretraining strategies that incorporate additional sources of information.
PNMTA~\cite{pnmta} proposes a meta-learning-based method that makes use of the existing non-cold-start users for pretraining to improve cold-start recommendation. 
P5~\cite{p5} and M6-Rec~\cite{m6rec} convert all data in recommender systems to natural language sequences and deploy language models for pretraining. 
Compared to these methods, our proposed framework is model-agnostic and only requires graph-structured data during pretraining, which are widely used in recommender systems~\cite{glrs_survey}, instead of depending on relatively rare visual or textual information.

\section{Preliminaries}

\begin{mydef}
\textbf{(Recommendation Task)}. 
Given two disjoint node sets, including a user set $\mathcal{U}$ and an item set $\mathcal{I}$, and the interactive edges, i.e., user-item edges $E_{\mathcal{U},\mathcal{I}}$, an interaction graph is defined as $\mathcal{G}=(V,E)$ where $V = \mathcal{U} \cup \mathcal{I}$. A personalized recommendation task  $t_{rec}$ for a user $u$ is to predict a ranking list of items $\{i_1, i_2, \dots, i_m\}$, with which this user has no interactions in the graph $\mathcal{G}$.
\end{mydef}
\begin{mydef}
\textbf{(Auxiliary Graph-based Pretext Task)}. 
Within an auxiliary pretext task set $T$, for a specific task $t\in T$, given an additional node property set $A$ or an additional edge set $E_{t}$ associated with the task-related node set $V_t\subseteq V$, an auxiliary graph-based pretext task is to predict the value of property $a_t\in A_t$ for an unlabeled node $v_t$, or the unobserved edge $e_t$ in the auxiliary graph $\mathcal{G}_t=(V_t,E_t)$.
\end{mydef}

\begin{mydef}
\textbf{(Task hypergraph)}. 
Given a pretext task $t$ which is $t_{rec}$ or an auxiliary pretext task $t\in T$, the task hypergraph is denoted as $\mathcal{H}_t=(V_t,\mathcal{E}_t)$, where $V_t$ are nodes related to the specific task in the interaction graph $\mathcal{G}$ and $\mathcal{E}_t$ are task-specific hyperedges. An incidence matrix $\mathbf{H}^{\mathbf{v}}_{t}\in\{0,1\}^{|V_t|\times|\mathcal{E}_t|}$ is used to represent connections among the related nodes.
\end{mydef}
In other words, we leverage hyperedges to represent associations among task-related nodes in each pretext task. 

\begin{mydef}
\textbf{(Hypergraph Encoder)}
The hypergraph encoder in our framework consists of hypergraph convolution layers, each of which is formulated as:
\begin{equation}\label{eq_encoder}
\begin{aligned}
\mathbf{E'}^{\mathbf{v}}_{t}&=(\mathbf{D}^{\mathbf{v}}_{t})^{-1}\mathbf{H}^{\mathbf{v}}_{t}\cdot\mathbf{E}^{\mathbf{\epsilon}}_{t},\\
&=(\mathbf{D}^{\mathbf{v}}_{t})^{-1}\mathbf{H}^{\mathbf{v}}_{t}\cdot(\mathbf{B}^{\mathbf{v}}_{t})^{-1}(\mathbf{H}^{\mathbf{v}}_{t})^\top\mathbf{E}^{\mathbf{v}}_{t},
\end{aligned}
\end{equation}
where $\mathbf{E}^{\mathbf{v}}_{t}$ and $\mathbf{E'}^{\mathbf{v}}_{t}$ are input and output embeddings of task-related node set $V_t$ for a certain pretext task $t$ in a hypergraph convolution layer, $\mathbf{H}^{\mathbf{v}}_{t}$ is the incidence matrix of the task hypergraph, $\mathbf{E}^{\mathbf{\epsilon}}_{t}$ denotes the hyperedge embeddings, $\mathbf{D}^{\mathbf{v}}_{t}$ is the degree matrix of hyperedges and $\mathbf{B}^{\mathbf{v}}_{t}$ is the degree matrix of nodes for normalization.  
\end{mydef}

\section{Methods}
In this section, we present the proposed UPRTH framework for unified pretraining. The illustration of UPRTH is demonstrated in Figure~\ref{fig:UPRTH}. We start by introducing all the embeddings to be pretrained in this framework. Thereafter, we define three types of pretext tasks and explain how to construct the task hypergraph corresponding to each type of tasks. Specifically, we adopt a TA layer to learn the relevance between each auxiliary task and the recommendation task, and then disciminatively integrate the prior knowledge from pretext tasks to the pretrained embeddings. Our framework can be easily extended to other graph-based tasks besides recommendation by replacing the target recommendation task. We leave this extension to the future work.
\subsection{Embedding Layer}
We maintain an embedding layer $\mathbf{E}\in \mathbb{R}^{d\times(|{\mathcal{U}}|+|\mathcal{I}|)}$, where $d$ is the feature dimension and columns represent all trainable embeddings for users and items during pretraining. We denote the initial user embedding as $\mathbf{E}^{\mathbf{u}}_{t_{rec}}$, and the initial item embedding as $\mathbf{E}^{\mathbf{i}}_{t_{rec}}$. By using the pretrained embeddings $\mathbf{E}^{\mathbf{u}}_{t_{rec}}$ and $\mathbf{E}^{\mathbf{i}}_{t_{rec}}$, prior knowledge learned from the pretraining phase can be leveraged when building a downstream recommendation model.

\begin{figure}[ht]
\centering
    \includegraphics[width=\linewidth]{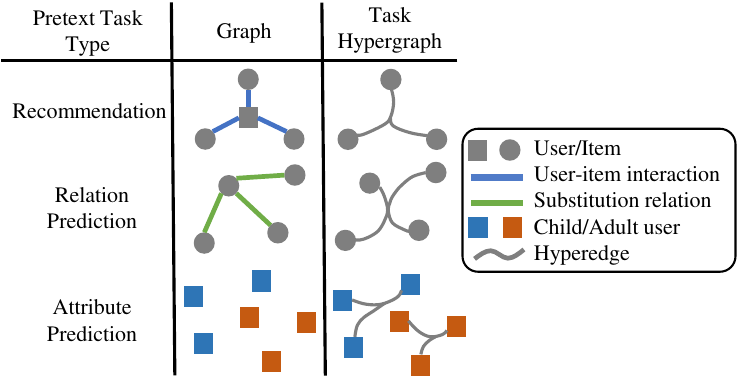}
  \caption{Construction of task hypergraphs.}
  \label{fig:construction}
\end{figure}

\subsection{Task Hypergraph Construction}
For the purpose of a unified architecture to handle various pretext tasks, we first generalize each pretext task to the format of a corresponding task hypergraph. A toy example is shown in Figure~\ref{fig:construction}. Compared to a graph where edges only connect two nodes, the advantage of a hypergraph where hyperedges can simultaneously connect multiple nodes lies in its ability to represent complex relationships. In multitask pretraining, the association among task-related nodes in each pretext task is diverse. Thus, we design a task hypergraph for each pretext task, in order to express these associations in different pretext tasks in a unified format of hyperedges for further hypergraph learning. 

Without loss of generality, we categorize all pretext tasks into three types: \textit{Recommendation}, \textit{Relation Prediction}, and \textit{Attribute Prediction}. Then we propose the construction of the hypergraph corresponding to each type as follows:
\begin{itemize}[leftmargin=*]
    \item \textbf{Recommendation}. Recommendation is to predict an unobserved relation between users and items, including purchasing, clicking, viewing or other user-item interactions. In this case, we let each type of node be the hyperedges to connect the other type of nodes. For example, if multiple users purchase an item, a hyperedge is constructed to connect them simultaneously. Two incident matrices $\mathbf{H}^{\mathbf{u}}_{t_{rec}}\in\{0,1\}^{|\mathcal{U}|\times|\mathcal{I}|}$ and $\mathbf{H}^{\mathbf{i}}_{t_{rec}}\in\{0,1\}^{|\mathcal{I}|\times|\mathcal{U}|}$ are used to represent the hyperedge connection in such tasks.
    \item \textbf{Relation Prediction}. The tasks belonging to this type are to predict an unobserved relation among nodes with the same type, such as whether a substitution relation exists between item nodes. In this type of task, we let the homogeneous relation itself be the hyperedge. For example, items that are substitutions with a target item are connected by a hyperedge. And the incidence matrix for this hypergraph is $\mathbf{H}^{\mathbf{i}}_{t}\in\{0,1\}^{|\mathcal{I}|\times|\mathcal{E}^{\mathbf{i}}_{t}|}$ where $|\mathcal{E}^{\mathbf{i}}_{t}|$ represents the number of substitution relations.
    \item \textbf{Attribute Prediction}. In this type of task, we let each type of attribute be the hyperedge to connect the nodes with such a type of attribute. For example, in a category classification task, items belonging to the same category are connected by one hyperedge. For the attributes whose values are continuous, such as ratings of items, we quantize those values to certain discrete magnitudes. The incidence matrix is denoted by $\mathbf{H}^{\mathbf{u}}_{t}\in\{0,1\}^{|\mathcal{U}|\times|A_t|}$ or $\mathbf{H}^{\mathbf{i}}_{t}\in\{0,1\}^{|\mathcal{I}|\times|A_t|}$, where $|A_t|$ is the number of possible values of the attribute to predict in the task $t$.
\end{itemize}
Both relation prediction and attribute prediction contain more fine-grained pretext tasks, such as friend recommendation and rating prediction.
For all three types of pretext tasks, hyperedges are leveraged to connect homogeneous nodes in our construction of task hypergraphs. In this way, users or items with specific commonalities corresponding to this task are connected by one hyperedge for a certain task. Henceforth, we can effectively capture collaborative signals among users and items from different pretext tasks through hypergraph learning.

\begin{figure*}[ht]
  \centering
    \includegraphics[width=\linewidth]{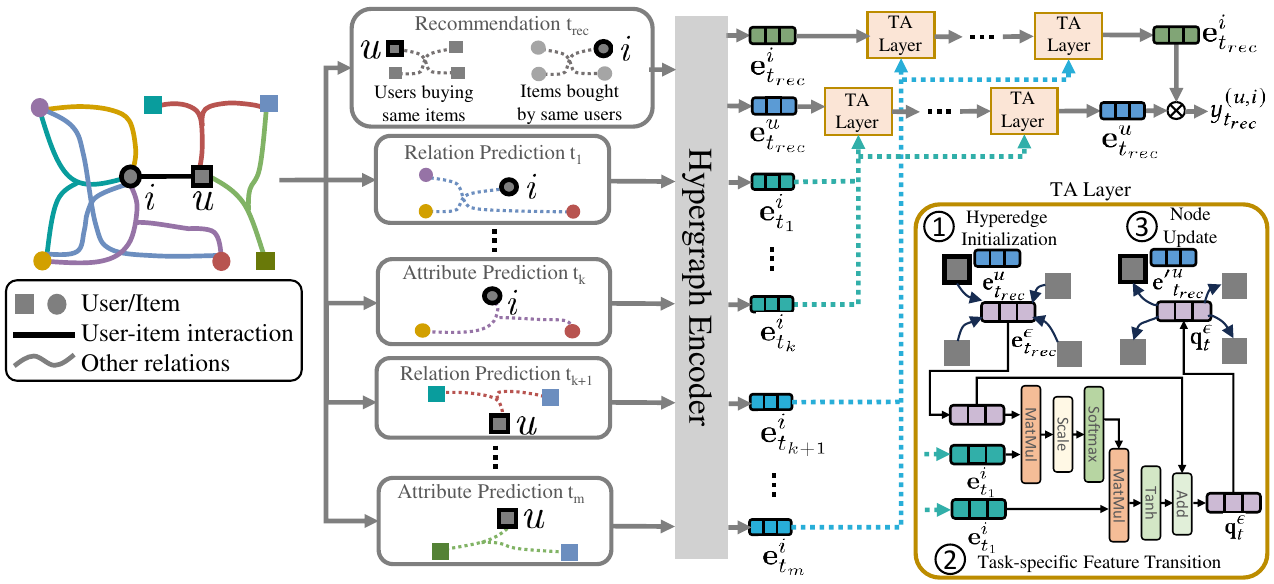}
  \caption{(a) The overall framework of UPRTH;
  First, we construct task hypergraphs according to different pretext tasks. Next, we apply hypergraph encoders to learn task-specific embeddings from task hypergraphs. At last, we use  TA layers to discriminatively transfer knowledge from auxiliary pretext tasks to the recommendation task; 
  (b) the illustration of the Transitional Attention (TA) layer.}
  \label{fig:UPRTH}
\end{figure*}

\subsection{Transitional Attention Layer}
To leverage multiple pretext tasks to improve the performance of the target recommendation task, it is important to adaptively learn the degrees of relevance between each auxiliary pretext task and the target recommendation task during pretraining. In this way, the model can identify which tasks provide significant insights that are crucial for the recommendation task. Hence, we propose a Transitional Attention (TA) Layer to ensure that the model allocates appropriate attention to each task based on its relevance, thereby enabling effective knowledge transfer during pretraining. The illustration is in Figure~\ref{fig:UPRTH}(b). Next, we present the calculation details of a TA layer that discriminatively transfers knowledge from item-related auxiliary pretext tasks to the recommendation task by pretraining user embeddings. For simplicity, we use $T$ to represent the set of item-related auxiliary tasks in this section. And we denote the output user embedding as $\mathbf{e'}^{u}_{t_{rec}}$ to distinguish it from the input user embedding $\mathbf{e}^{u}_{t_{rec}}$ only for this section.

\subsubsection{Hyperedge Initialization}
Following the construction of the task hypergraph for the recommendation task, in Figure~\ref{fig:UPRTH}(b)\textcircled{1}, we use item hyperedges to connect users and initialize those hyperedges by aggregating users' preferences as below:
\begin{equation}\label{eq_initialization}
    \mathbf{e}^{\epsilon}_{t_{rec}} = \frac{1}{|\mathcal{N}_\epsilon|}\sum_{u^*\in\mathcal{N}_\epsilon}\mathbf{e}^{u^*}_{t_{rec}},
\end{equation}
where $\mathbf{e}^{\epsilon}_{t_{rec}}$ denotes temporary embedding of the hyperedge $\epsilon$, $\mathcal{N}_\epsilon$ is the set of user connected by this hyperedge, and $\mathbf{e}^{u^*}_{t_{rec}}$ is the embedding of those users as input of the $l$-th TA layer.
$\mathbf{e}^{\epsilon}_{t_{rec}}$ represents the preferences of users who are connected by this hyperegde.  

\subsubsection{Task-specific Feature Transition}
In this step, we first calculate the attention between the users' preferences and the task-specific item features from each auxiliary pretext task. And then we generate task-aware hyperedge embeddings accordingly. Inspired by self-attention mechanism~\cite{transformer}, 
the attention is calculated based on the hyperedge embedding and the task-specific item embedding for item $i$ from each auxiliary pretext task:
\begin{equation}\label{eq_attention}
\begin{aligned}
    \mathbf{a}^{\epsilon}_{t} &= \sigma(\sum_{t \in T}\alpha^{i}_{t}\cdot\mathbf{e}^{i}_{t}),\\
    \alpha^{i}_{t} &= \frac{\text{exp}(\mathbf{e}^{\epsilon}_{t_{rec}}\cdot\mathbf{e}^{i}_{t}/\sqrt{d})}
    {\sum_{t^* \in T}\text{exp}(\mathbf{e}^{\epsilon}_{t_{rec}}\cdot\mathbf{e}^{i}_{t^*}/\sqrt{d})}, 
\end{aligned}
\end{equation}
where $\mathbf{a}^{\epsilon}_{t}$ is the task-aware hyperedge embedding. Embeddings $\mathbf{e}^{i}_{t}$ and $\mathbf{e}^{i}_{t^*}$ are embeddings of item node $i$ in auxiliary pretext tasks $t$ and $t^*$, respectively. 
Attention $\alpha^{i}_{t}$ represents the attention of the connected users, who are interested in this item, on its certain feature reflected by each task.
And $\sigma$ is \textit{tanh()} as the activation function. 
Next, we fuse $\mathbf{a}^{\epsilon}_{t}$ with $\mathbf{e}^{\epsilon}_{t_{rec}}$ by addition:
\begin{equation}\label{eq_transition}
    \mathbf{q}^{\epsilon}_{t} = \mathbf{e}^{\epsilon}_{t_{rec}} + \gamma\mathbf{a}^{\epsilon}_{t},
\end{equation}
where $\mathbf{q}^{\epsilon}_{t}$ is the fused hyperedge embedding, and $\gamma$ is a scalar hyper-parameter to control the transition intensity. This step is demonstrated in \textcircled{2} of Figure~\ref{fig:UPRTH}(b).  

\subsubsection{Node update} We update the user embedding by aggregating the fused embeddings from all its connected hyperedges.
For a target user $u$, the aggregation step is formulated as follows:
\begin{equation}\label{eq_update}
    \mathbf{e'}^{u}_{t_{rec}}= \frac{1}{|\mathcal{N}_u|}\sum_{\epsilon\in\mathcal{N}_u}\mathbf{q}^{\epsilon}_{t},
\end{equation}
where $\mathbf{e'}^{u_{out}}_{t_{rec}}$ denotes the output user embedding, $\mathcal{N}_u$ is the set of hyperedges connected to this user, and $\mathbf{q}^{\epsilon}_{t}$ is the fused embedding of its connected hyperedges from Eq.~(\ref{eq_transition}).
This step is demontrated in Figure~\ref{fig:UPRTH}(b)\textcircled{3}. 

\subsubsection{TA Layer in Matrix Form}
To offer a holistic view, we formulate the matrix form of the transitional attention layer (equivalent to Eqs.~(\ref{eq_initialization})-(\ref{eq_update})) as:
\begin{equation}\label{ta}
\begin{aligned}
    \mathbf{E'}^{\mathbf{u}}_{t_{rec}}
    &= \text{TA}(\mathbf{E}^{\mathbf{u}}_{t_{rec}},\mathbf{H}^{\mathbf{u}}_{t_{rec}},\mathbf{E}^{\mathbf{i}}_{t})
    =(\mathbf{D}^{\mathbf{u}}_{t_{rec}})^{-1}\mathbf{H}^{\mathbf{u}}_{t_{rec}}\cdot\mathbf{Q}^{\mathbf{\epsilon}}_{t}\\ 
    \mathbf{Q}^{\mathbf{\epsilon}}_{t}&= (\mathbf{E}^{\mathbf{\epsilon}}_{t_{rec}} + \gamma\cdot\sigma(\text{Softmax}(\frac{\mathbf{E}^{\mathbf{\epsilon}}_{t_{rec}}\mathbf{E}^{\mathbf{i}}_{t}}{\sqrt{d}})\mathbf{E}^{\mathbf{i}}_{t})),\\
    \mathbf{E}^{\mathbf{\epsilon}}_{t_{rec}} &= (\mathbf{B}^{\mathbf{u}}_{t_{rec}})^{-1}(\mathbf{H}^{\mathbf{u}}_{t_{rec}})^\top\mathbf{E}^{\mathbf{u}}_{t_{rec}},
\end{aligned}
\end{equation}
where $\mathbf{E}^{\mathbf{u}}_{t_{rec}}$ is the input embedding from $l$-th layer, $\mathbf{H}^{\mathbf{u}}_{t_{rec}}$ is the incidence matrix, $\mathbf{E}^{\mathbf{i}}_{t}$ denotes task-specific item features from auxiliary pretext tasks, and $\mathbf{E'}^{\mathbf{u}}_{t_{rec}}$ is the output. $\mathbf{D}^{\mathbf{u}}_{t_{rec}}$ is the degree matrix of nodes and $\mathbf{B}^{\mathbf{u}}_{t_{rec}}$ is degree matrix of hyperedges. First, we aggregate preferences of users to the items they are interested in. Next, we calculate the attention to reflect the relevance between the user preferences and the item features related to each task. At last, the task-aware user preferences are aggregated back to each user to support recommendation.

\subsection{Prediction and Optimization}
We conduct prediction and optimization for different tasks depending on their types. For bipartite relation prediction, i.e., the recommendation task, we calculate the ranking score ${y}^{(u,i)}_{t_{rec}}$ of the user-item pair $(u,i)$ by the inner product: 
\begin{equation}
{y}^{(u,i)}_{t_{rec}} = \mathbf{e}^{u}_{t_{rec}}\cdot \mathbf{e}^{i}_{t_{rec}},
\end{equation}
where $\mathbf{e}^{u}_{t_{rec}}$ and $\mathbf{e}^{i}_{t_{rec}}$ are user and item embeddings output by the last TA layer. We apply the alignment loss~\cite{auloss} to optimize the recommendation:
\begin{equation}
\mathcal{L}_{rec}=\sum\limits_{(u,i)\in \mathcal{D}_r} ||\mathbf{e}^{u}_{t_{rec}}-\mathbf{e}^{i}_{t_{rec}}||^{2},
\end{equation}
where $\mathcal{D}_r=\{(u,i)|i\in \mathcal{G}^{+}_{u}$ is the user-item interaction data for pretraining. 

For homogenous relation and node attribute prediction, since relations and attributes are represented by hyperedges, we use the inner product between the node embedding and the corresponding hyperedge embedding as the prediction score:
\begin{equation}
{y}^{(v,\epsilon)}_{t} = \mathbf{e}_{(v,t)}\cdot \mathbf{e}^{\epsilon}_{t},
\end{equation}
where $\mathbf{e}_{(v,t)}$ and $\mathbf{e}^{\epsilon}_{t}$ are the task-related node embedding and the hyperedge embedding for a specific auxiliary pretext task $t$, as calculated in Eq.~(\ref{eq_encoder}).
Then we adopt the pairwise Bayesian Personalized Ranking (BPR) loss~\cite{bprloss} to optimize the prediction:
\begin{equation}
\mathcal{L}_{t}=\sum\limits_{(v,\epsilon,\epsilon')\in \mathcal{D}_a} -\log\sigma(\hat{y}^{(v,\epsilon)}_{t}-\hat{y}^{(v,\epsilon')}_{t},
\end{equation}
where $\mathcal{D}_a=\{(v,\epsilon,\epsilon')|\epsilon\in \mathcal{G}^{+}_{a}, \epsilon'\in \mathcal{G}_{a}\backslash\mathcal{G}^{+}_{a}\}$ is the pretraining data for each auxiliary pretext task. 

Finally, we jointly optimize the recommendation pretext task and the auxiliary pretext tasks as follows:
\begin{equation}
\mathcal{L}= \beta\cdot\mathcal{L}_{rec} + (1-\beta)\cdot\sum\limits_{t\in T}\mathcal{L}_{t} + \lambda_\Theta\|\Theta\|^2_2,
\end{equation}
where $\Theta$ is all trainable parameters in the framework, which is regularized by $\lambda_\Theta$, and $\beta$ is a coefficient to balance the loss of recommendation and auxiliary tasks. 
Adam~\cite{adam14} is chosen as the optimizer.

The trainable parameters of UPRTH are only the initial user and item embeddings, i.e., $\Theta=\{\mathbf{E}^{\mathbf{u}}_{t_{rec}}\cup\mathbf{E}^{\mathbf{i}}_{t_{rec}}\}$. In other words, the complexity of encoding task hypergraphs is $\mathcal{O}(|V|^2)$ as the standard matrix factorization~\cite{lightgcn20}. Without additional trainable parameters, the proposed TA layer with the softmax dot-product attention takes $\mathcal{O}(|V|^2)$. Thus, the model complexity is $\mathcal{O}(|V|^2)$, which is as efficient as the matrix factorization.

\section{Experiment}
\subsection{Experimental Setup}
\subsubsection{Datasets} 
We conduct experiments on three real-world datasets: XMarket-CN,  XMarket-MX, and Steam. 
XMarket\footnote{\url{https://xmrec.github.io/}} is a publicly available large-scale dataset obtained from Amazon. We use the data from China and Mexico for our experiments. For these two datasets, XMarket-CN and XMarket-MX, item categories and average rating of items are predicted as node attributes in auxiliary pretext tasks. And items bought together are regarded as items in complementary relations, which are predicted as homogeneous edges. In addition, substitute relations between items compared together are predicted as homogeneous edges for XMarket-MX.
Steam~\cite{steam} includes users' transaction records on the Steam online game store. Categories, publishers of the games, and groups of users are used for attribute prediction in auxiliary pretext tasks. Friendships among users are predicted as homogeneous edges.
The statistics of three datasets are shown in Table \ref{tab:dataset}. 
For Steam and XMarket-CN, we randomly select $80\%$ of user-item interactions for training in both pretraining and finetuning and the remaining $20\%$ for testing in finetuning. 
For XMarket-MX, the split ratio is $70\%$ for training and $30\%$ for testing. 

\begin{table}\caption{The statistics of datasets.}\label{tab:dataset}
\resizebox{0.47\textwidth}{!}{
\begin{tabular}{l|l|l|l}
\hline
\hline
Dataset               & XMarket-CN & XMarket-MX & Steam  \\ \hline
\# users              & 18,806     & 221,890    & 50,292 \\
\# items              & 5,937      & 35,235     & 1,809  \\
\# user-item edges    & 23,065     & 305,569    & 65,379 \\
\# auxiliary tasks    & 3          & 4          & 4     \\
\# node attributes    & 11,874         & 70,470         & 215,285      \\
\# homogeneous edges& 318      & 8,212      & 6,398      \\ \hline
\end{tabular}
}
\end{table}

\subsubsection{Baselines}
As UPRTH is a pretraining framework, we compare the performances of graph-based recommendation models with and without using user and item embeddings pretrained by UPRTH, including LGCN~\cite{lightgcn20}, GCN~\cite{GCN16}, DirectAU~\cite{auloss}, UltraGCN~\cite{ultragcn}, HGNN~\cite{hnn}, HCCF~\cite{hccf} and DHCF~\cite{dhcf}.
We also compare UPRTH with the following pretraining baselines:
\begin{itemize}[leftmargin=*]
    \item \textbf{GCC}~\cite{gcc}. This self-supervised pretraining framework leverages contrastive learning with random walks to capture structure information in graphs.
    \item \textbf{SGL-PreRec}. In SGL-Pre~\cite{sgl}, edge drop as data augmentation is used for contrastive learning in pretraining. To further differentiate it from GCC, we jointly pretrain the model via both the contrastive learning task loss and the recommendation task.
    \item \textbf{AttriMask}~\cite{weihua20}. This method applies GNN to obtained node embeddings and then adds a linear model to predict masked attributes during pretraining.
\end{itemize}
In finetuning of pretraining baselines and UPRTH, we deploy one hypergraph convolution layer to encode user and item embeddings on the hypergraph of the recommendation task, and use BPR loss for optimization. Our source code and the three datasets are released online \footnote{https://github.com/mdyfrank/UPRTH}.

\begin{table*}[!ht]\caption{Performance comparison on three datasets.}\label{tab_1}
\begin{tabular}{l|cccc|cccc|cccc}
\hline
Dataset    & \multicolumn{4}{c|}{XMarket-CN}                                       & \multicolumn{4}{c|}{XMarket-MX}                                       & \multicolumn{4}{c}{Steam}                                             \\ \hline
Metric     & R@10            & R@20            & N@10            & N@20            & R@10            & R@20            & N@10            & N@20            & R@10            & R@20            & N@10            & N@20            \\ \hline
LightGCN   & 0.0036          & 0.0068          & 0.0015          & 0.0024          & 0.0123          & 0.0181          & 0.0073          & 0.0089          & 0.0715          & 0.1075          & 0.0358          & 0.0448          \\
DirectAU   & 0.0030          & 0.0040          & 0.0014          & 0.0017          & 0.0098          & 0.0160          & 0.0055          & 0.0072          & 0.0574          & 0.0860          & 0.0290          & 0.0362          \\
UltraGCN   & 0.0031          & 0.0053          & 0.0017          & 0.0023          & 0.0017          & 0.0026          & 0.0009          & 0.0011          & 0.0416          & 0.0787          & 0.0206          & 0.0299          \\
HGNN       & 0.0061          & 0.0074          & 0.0029          & 0.0033          & 0.0095          & 0.0154          & 0.0046          & 0.0062          & 0.0780          & 0.1104          & 0.0418          & 0.0502          \\
HCCF       & 0.0023          & 0.0038          & 0.0008          & 0.0012          & 0.0077          & 0.0122          & 0.0043          & 0.0055          & 0.0976          & 0.1453          & 0.0489          & 0.0609          \\
DHCF       & 0.0033          & 0.0047          & 0.0022          & 0.0026          & 0.0139          & 0.0230          & 0.0073          & 0.0097          & 0.1214          & 0.1959          & 0.0566          & 0.0753          \\ \hline
GCC        & 0.0112          & 0.0177          & 0.0068          & 0.0086          & 0.0229          & 0.0318          & 0.0146          & 0.0169          & 0.1147          & 0.1833          & 0.0565          & 0.0738          \\
SGL-PreRec & 0.0036          & 0.0079          & 0.0014          & 0.0025          & 0.0102          & 0.0153          & 0.0058          & 0.0071          & 0.1120          & 0.1796          & 0.0537          & 0.0708          \\
AttriMask  & {\ul 0.0387}    & {\ul 0.0693}    & {\ul 0.0179}    & {\ul 0.0261}    & {\ul 0.0500}    & {\ul 0.0748}    & {\ul 0.0261}    & {\ul 0.0327}    & {\ul 0.1398}    & {\ul 0.2247}    & {\ul 0.0714}    & {\ul 0.0929}    \\ \hline
UPRTH       & \textbf{0.0801} & \textbf{0.1093} & \textbf{0.0428} & \textbf{0.0506} & \textbf{0.0822} & \textbf{0.0966} & \textbf{0.0574} & \textbf{0.0612} & \textbf{0.1721} & \textbf{0.2704} & \textbf{0.0841} & \textbf{0.1088} \\ \hline
Improv.    & 106.98\%        & 57.72\%         & 139.11\%        & 93.87\%         & 64.40\%         & 29.14\%         & 119.92\%        & 87.16\%         & 13.07\%         & 23.10\%         & 17.79\%         & 17.12\%         \\ \hline
\end{tabular}
\end{table*}


\subsubsection{Evaluation Metrics}
We evaluate the pretraining framework by ranking the test items with all non-interacted users during finetuning. Recall$@ \{10,20\}$ and NDCG$@\{10,20\}$ are adopted as evaluation metrics. 

\subsection{Overall Performance Comparison}
We show overall comparison results in Table~\ref{tab_1}. The best results are in boldface. We summarize the following key observations:
\begin{itemize}[leftmargin=*]
    \item The proposed UPRTH framework achieves the best results and outperforms all the baseline pretraining frameworks in the three datasets. 
    We hypothesize these large stable gains result from the abundant and balanced contextual information learned from unified pretext tasks.
    \item Pretraining frameworks achieves better performance than recommendation models without pretraining in most cases. However, pretraining based on only user-item interactions sometimes degrades the performance. Besides sparse data providing limited prior knowledge during pretraining, We believe that overfitting caused by optimizing the same recommendation objective in both pretraining and finetuning also leads to such worse performance.
    \item Among the three baseline pretraining frameworks, AttriMask which additionally considers attributes in pretraining is better than GCC and SGL-PreRec in all datasets. This justifies the necessity of capturing contextual cues besides user-item interactions in pretraining. Compared with it, UPRTH realizes unification across various pretext tasks so it obtains the capacity to transfer knowledge from a wider range of tasks in pretraining, which is a more comprehensive framework.
\end{itemize}

\begin{table}[ht]\caption{Performance of recommendation models with and without using embeddings pretrained by UPRTH.}\label{tab_2}
\resizebox{0.48\textwidth}{!}{
\begin{tabular}{l|cccccc}
\hline
Dataset                  & \multicolumn{2}{c}{XMarket-CN} & \multicolumn{2}{c}{XMarket-MX} & \multicolumn{2}{c}{Steam} \\ \hline
Metric                   & R@10           & N@10          & R@10           & N@10          & R@10        & N@10        \\ \hline
LightGCN                 & 0.0036         & 0.0015        & 0.0123         & 0.0073        & 0.0715      & 0.0358      \\
+UPRTH & 0.0679         & 0.0393        & 0.0680         & 0.0407        & 0.1522      & 0.0740      \\ \hline
DirectAU                 & 0.0030         & 0.0014        & 0.0098         & 0.0055        & 0.0574      & 0.0290      \\
+UPRTH & 0.0612         & 0.0352        & 0.0714         & 0.0490        & 0.1473      & 0.0727      \\ \hline
UltraGCN                 & 0.0031         & 0.0017        & 0.0017         & 0.0009        & 0.0416      & 0.0206      \\
+UPRTH & 0.0607         & 0.0412        & 0.0644         & 0.0383        & 0.1049      & 0.0460      \\ \hline
HGNN                     & 0.0061         & 0.0029        & 0.0095         & 0.0046        & 0.0780      & 0.0418      \\
+UPRTH & 0.0510         & 0.0264        & 0.0411         & 0.0195        & 0.0951      & 0.0475      \\ \hline
HCCF                     & 0.0023         & 0.0008        & 0.0077         & 0.0043        & 0.0976      & 0.0489      \\
+UPRTH & 0.0341         & 0.0223        & 0.0376         & 0.0234        & 0.1143      & 0.0556      \\ \hline
DHCF                     & 0.0033         & 0.0022        & 0.0139         & 0.0073        & 0.1214      & 0.0566      \\
+UPRTH & 0.0562         & 0.0322        & 0.0382         & 0.0220        & 0.1408      & 0.0625      \\ \hline
\end{tabular}
}
\end{table}

Next, we show the performance of baseline recommendation models with and without leveraging pretrained embeddings from UPRTH in Table~\ref{tab_2}. The results indicate that all recommendation models benefit from embeddings pretrained by UPRTH, especially in the XMarket-CN and XMarket-MX datasets with sparser user-item interactions. It indicates our framework with high generalization can adapt well to various downstream recommendation models, showing robustness and flexibility in its pretrained embeddings. The benefit on sparse datasets reflects that unified multitask pretraining efficiently addresses the issue of data sparsity in the recommendation task.
\subsection{Ablation Study}
\subsubsection{Unification across Pretext Tasks.}
\begin{figure}[ht]
    \begin{subfigure}{0.155\textwidth}
    \includegraphics[width=\textwidth]{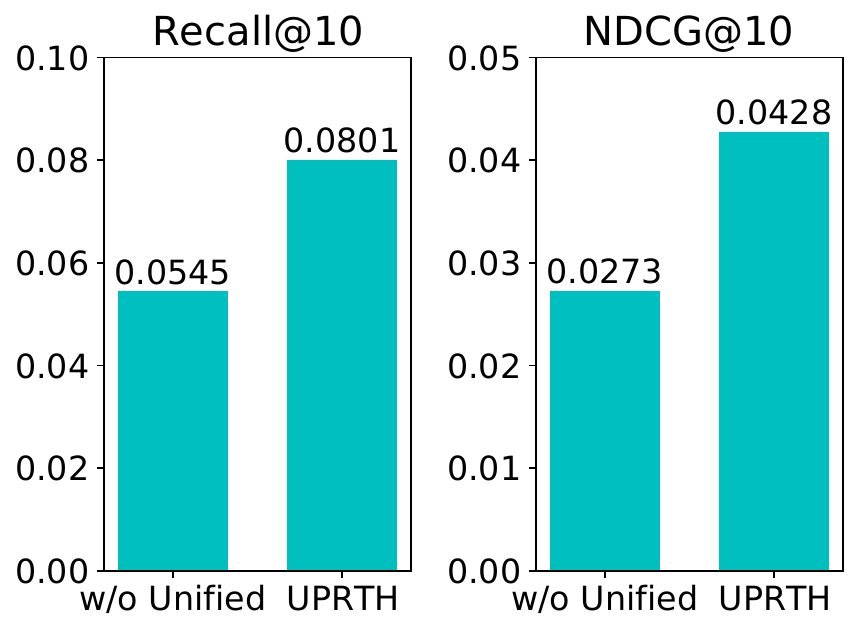}
    \caption{XMarket-CN}
    \end{subfigure}
    \hfill
    \begin{subfigure}{0.155\textwidth}
    \includegraphics[width=\textwidth]{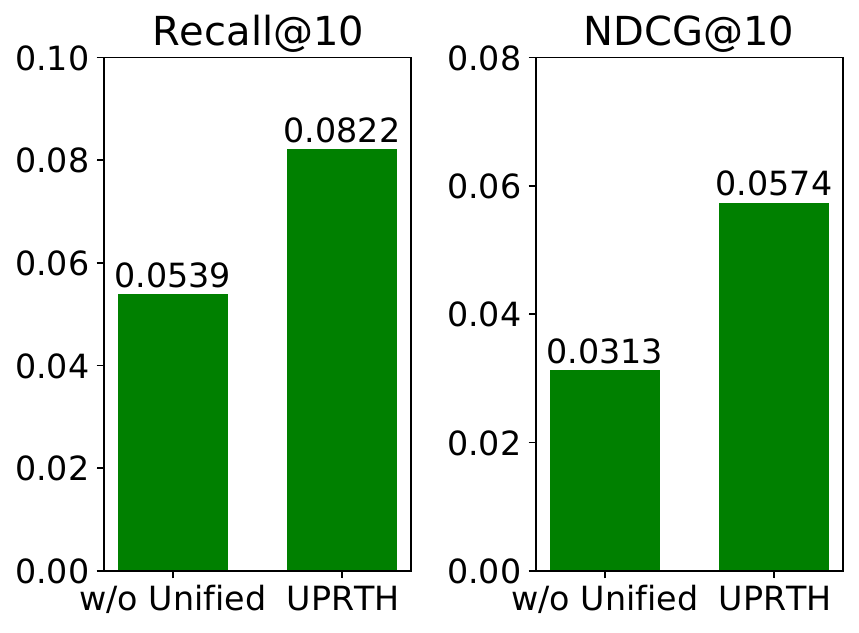}
    \caption{XMarket-MX}
    \end{subfigure}
    \hfill
    \begin{subfigure}{0.155\textwidth}
    \includegraphics[width=\textwidth]{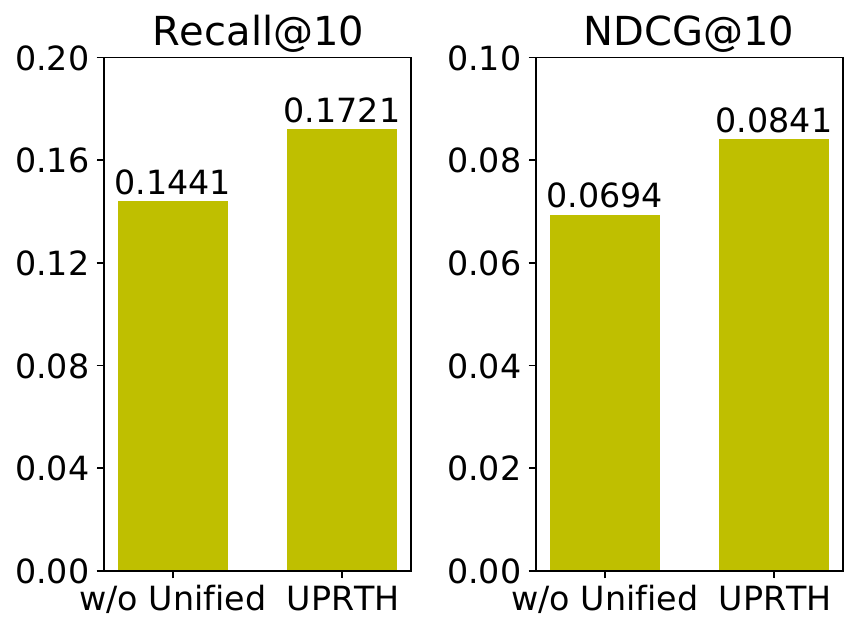}
    \caption{Steam}
    \end{subfigure}
        \caption{Performance of UPRTH with and without unified pretraining.}
  \label{fig:UnifyAblation}
\end{figure}

We compare the performance of UPRTH and its variant without unified pretext tasks on the three datasets in Figure~\ref{fig:UnifyAblation}. In the variant Without unified pretext tasks, relations are still predicted as hyperedges in relation prediction tasks, but attributes are predicted via applying one linear layer followed by a softmax activation function on node embeddings output by hypergraph encoders. We can observe that the performance is consistently improved by a unified pretraining framework on all three datasets. We attribute this improvement to the complete transfer of prior knowledge from our unified multitask pretraining.
Given that no linear transformation is included in hypergraph encoders, and the prediction of hyperedges is directly calculated based on the node embeddings, prior knowledge from a pretext task is completely preserved in node embeddings and transferred to the downstream recommendation task. Thus, we believe UPRTH is a concise and efficient design to unify pretext tasks.

\subsubsection{Transitiona Attention Layer Settings.}
To verify the effectiveness of TA, we demonstrate the performance of UPRTH and three other variants in Table~\ref{tab:ta}: (a) UH w/o TA. In this setting, embeddings output by hypergraph encoders are directly used to calculate ranking scores of user-item pairs; (b) UH-sum. Instead of calculating the attention of the hyperedge embedding on node embeddings from different auxiliary tasks, the sum of the hyperedge embedding and those node embeddings are used as the fused hyperedge embedding for further node update; (c) UH-concat. Same as UPRTH-sum, but we concatenate node embeddings from different auxiliary tasks and transform it to an embedding whose dimension is the same as the hyperedge embedding by a linear layer, and then add it to the hyperedge embedding as the fused hyperedge embedding.

We observe that UPRTH performs the best on three datasets. This justifies the superiority of UPRTH with TA layers which discriminatively integrates information from auxiliary tasks to recommendation. The poor performance of UH-concat verifies the negative impact brought by the additional linear transformation in pretraining. UH-sum performs better than UH w/o TA in three datasets, which indicates the necessity of considering contextual information from other tasks in the pretext recommendation task. However, UH-sum ignores the different relevance between each auxiliary task and recommendation, so it is worse than UPRTH.
\begin{table}[ht]\caption{Ablation study on Transitional Attention Layer.}\label{tab:ta}
\resizebox{0.48\textwidth}{!}{%
\begin{tabular}{l|cccccc}
\hline
Dataset & \multicolumn{2}{c}{XMarket-CN}      & \multicolumn{2}{c}{XMarket-MX}     & \multicolumn{2}{c}{Steam}         \\ \hline
Metric    & R@10   & N@10   & R@10   & N@10   & R@10   & N@10   \\ \hline
UH w/o TA  & 0.0744 & 0.0411 & 0.0798 & 0.0548 & 0.1645 &0.0795 \\
UH-sum  & 0.0770 & 0.0413 & 0.0818 & 0.0547 & 0.1690 & 0.0825 \\
UH-concat & 0.0800 & 0.0424 & 0.0796 & 0.0470 & 0.1235 & 0.0599 \\
\hline
UPRTH    & \textbf{0.0801} & \textbf{0.0428} & \textbf{0.0822} & \textbf{0.0574} & \textbf{0.1721} & \textbf{0.0841} \\ \hline
\end{tabular}%
}
\end{table}

\subsection{Pretraining and Finetuning Loss Analysis}
As aforementioned, UPRTH applies alignment loss in pretraining and BPR loss in finetuning to optimize the recommendation task. We further investigate different combinations of loss for pretraining and finetuning stages. In addition, we try AU loss~\cite{auloss} which optimizes both alignment and uniformity, and BPR$_{\text{pos}}$ loss which only optimize for positive instead of both positive and negative pairs in BPR loss. The performances are shown in Table~\ref{tab:loss}. We observe that using two different loss functions for pretraining and finetuning improves performance in most cases, since the same loss functions increase the risk of overfitting. Moreover, the results show that employing BPR loss in finetuning is better than in pretraining. We believe that exposing different pretext tasks to different objective functions results in more robust pretrained representations for downstream tasks.

\begin{table}[ht]\caption{Performance of UPRTH w.r.t. objective functions in pretraining and finetuning.
From top to bottom: Same loss for both stages, BPR for pretraining, and BPR for finetuning.}\label{tab:loss}
\resizebox{0.48\textwidth}{!}{
\begin{tabular}{ll|cccccc}
\hline
\multicolumn{2}{l|}{Dataset}       & \multicolumn{2}{c}{XMarket-CN}             & \multicolumn{2}{c}{XMarket-MX}             & \multicolumn{2}{c}{Steam}                  \\ \hline
\multicolumn{2}{l|}{$(\mathcal{L}_{\text{Pretrain}}$, $\mathcal{L}_{\text{Finetune}})$}         & R@10            & \multicolumn{1}{c}{N@10} & R@10            & \multicolumn{1}{c}{N@10} & R@10            & \multicolumn{1}{c}{N@10} \\ \hline
\multicolumn{2}{l|}{(BPR, BPR)}       & 0.0527          & 0.0284                   & 0.0707          & {\ul 0.0439}             & 0.1053          & 0.0527                   \\
\multicolumn{2}{l|}{(BPR$_{\text{pos}}$, BPR$_{\text{pos}}$)} & 0.0231          & 0.0114                   & 0.0632          & 0.0354                   & 0.1106          & 0.0564                   \\
\multicolumn{2}{l|}{(AU, AU)}         & 0.0460          & 0.0238                   & 0.0159          & 0.0084                   & 0.1152          & 0.0568                   \\
\multicolumn{2}{l|}{(Align., Align.)} & 0.0373          & 0.0237                   & 0.0110          & 0.0059                   & 0.0180          & 0.0069                   \\ \hline
\multicolumn{2}{l|}{(BPR, BPR$_{\text{pos}}$)}    & 0.0519          & 0.0283                   & 0.0633          & 0.0384                   & 0.1205          & 0.0601                   \\
\multicolumn{2}{l|}{(BPR, AU)}        & 0.0574          & 0.0284                   & 0.0643          & 0.0407                   & 0.1158          & 0.0575                   \\
\multicolumn{2}{l|}{(BPR, Align.)}    & 0.0575          & 0.0281                   & 0.0317          & 0.0125                   & 0.1062          & 0.0543                   \\ \hline
\multicolumn{2}{l|}{(BPR$_{\text{pos}}$, BPR)}    & {\ul 0.0623}    & 0.0301                   & 0.0704          & 0.0419                   & {\ul 0.1338}    & {\ul 0.0669}             \\
\multicolumn{2}{l|}{(AU, BPR)}        & 0.0623          & {\ul 0.0346}             & {\ul 0.0712}    & 0.0418                   & 0.1075          & 0.0545                   \\
\multicolumn{2}{l|}{(Align., BPR)} & \textbf{0.0801} & \textbf{0.0427}          & \textbf{0.0822} & \textbf{0.0579}          & \textbf{0.1721} & \textbf{0.0795}          \\ \hline
\end{tabular}%
}
\end{table}

\subsection{Cold-start Recommendation}
\begin{figure}[t]
    \begin{subfigure}{0.5\textwidth}
    \centering
    \includegraphics[width=\textwidth]{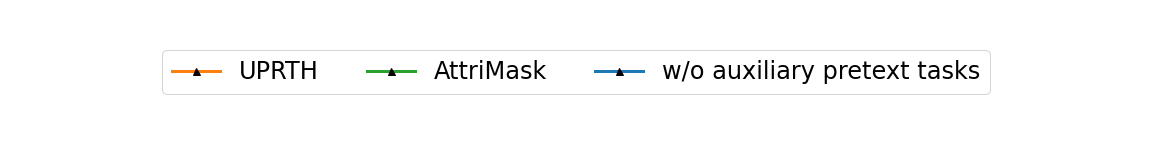}
    \end{subfigure}
    \vspace{-8mm}

    \begin{subfigure}{0.16\textwidth}
    \includegraphics[width=\textwidth]{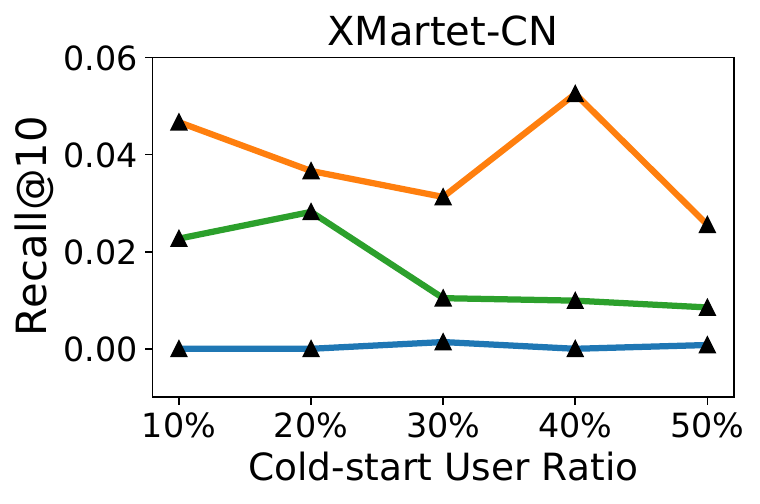}
    \end{subfigure}
    \hspace{-15mm}
    \hfill
    \begin{subfigure}{0.16\textwidth}
    \includegraphics[width=\textwidth]{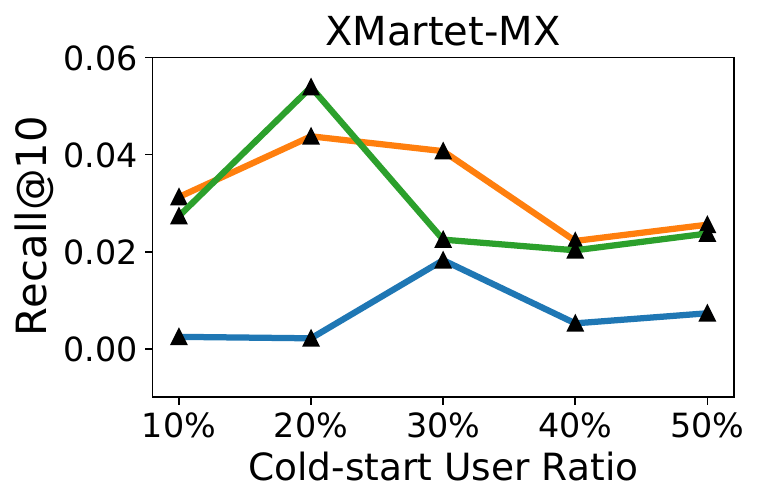}
    \end{subfigure}
    \hspace{-15mm}
    \hfill
    \begin{subfigure}{0.16\textwidth}
    \includegraphics[width=\textwidth]{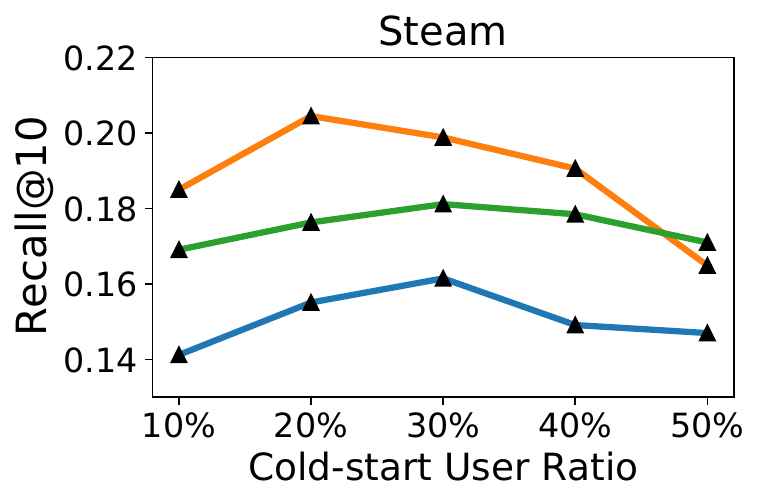}
    \end{subfigure}
  \caption{Performance on inductive users.}\label{fig:inductive_user}
\end{figure}

As an inductive framework, UPRTH has the capability of handling the cold-start problem by leveraging contextual information from auxiliary tasks.
We conduct an analysis regarding the ability of UPRTH to tackle users who have no historical user-item interactions in both pretraining and finetuning. We randomly select a certain ratio of user nodes as cold-start users and remove all user-item edges connected to them. The embeddings of cold-start users are obtained by feeding them with the removed edges to the downstream encoder during only inference. For comparison, we report testing recall of UPRTH, AttriMask and UPRTH without auxiliary pretext tasks. The results are shown in Figure~\ref{fig:inductive_user}. 

We observe that UPRTH and AttriMask perform better than UPRTH without auxiliary tasks, which verifies the significance of contextual information from multitask pretraining in cold-start recommendation. UPRTH outperforms AttriMask in most cases since more diverse contextual cues are captured for recommending items to cold-start users by unified multitask pretraining. Additionally, we observe that sometimes the performance on cold-start users is even better than the average performance on all users shown in Table~\ref{tab_1}, such as cold-start user ratio $r=20\%$ in Steam dataset. We hypothesize that when cold-start users are pretrained with only social relations, such as user-user and user-group hyperedges, UPRTH can well leverage user similarity for recommendation without the noise in the user-item interaction information.

\subsection{Performance Curves Analysis}
To verify that the pretrained embeddings with prior knowledge transferred from various pretext tasks can benefit downstream recommendation, we conduct analyses of testing performance curves with and without embeddings pretrained by UPRTH. For the downstream model, we choose the one-layer hypergraph convolution encoder used in UPRTH, denoted as HyperConv, with BPR loss for recommendation. Performances of SGL-PreRec, GCC and AttriMask are shown for comparison. The result indicates that UPRTH and AttriMask, which leverage contextual information besides user-item interactions from pretraining, outperform the other baselines. Compared with AttriMask, UPRTH discriminatively learns from various pretext tasks with a unified multitask framework, leading to more comprehensive and informative pretrained embeddings for downstream recommendation.

\begin{figure}[ht]\label{fig:training_curve}

    \centering
    \includegraphics[width=0.43\textwidth]{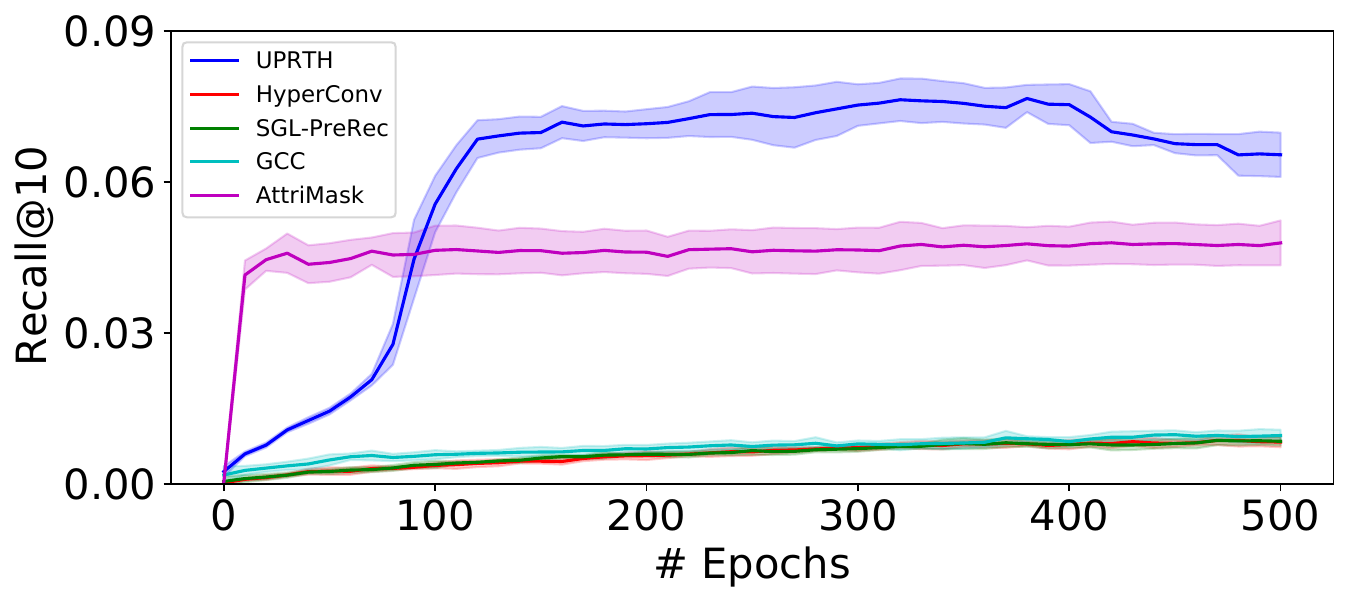}
    

  \caption{Testing recall w.r.t. the number of training epochs during finetuning in XMarket-MX dataset.}
\end{figure}

\section{Conclusion}
In this paper, we propose a novel pretraining framework UPRTH for recommendation. 
UPRTH leverages task hypergraphs to express complex relations in various pretext tasks, and applies hypergraph learning to capture prior knowledge from those unified pretext tasks. A TA layer is devised to adaptively learn the relevance between each auxiliary task and recommendation during pretraining.
We conduct extensive experiments and detailed
analyses on three datasets to verify the effectiveness of UPRTH. 
In the future, we may explore how to extend our framework for not only recommendation but also other graph-based downstream tasks.

\section*{ACKNOWLEDGEMENTS}
This work is supported by National Key R\&D Program of China through grant 2022YFB3104700, NSFC through grants 62322202, U21B2027, 62002007, 61972186 and 62266028, Beijing Natural Science Foundation through grant 4222030, S\&T Program of Hebei through grant 21340301D, Yunnan Provincial Major Science and Technology Special Plan Projects through grants 202302AD080003, 202202AD080003 and 202303AP140008, General Projects of  Basic Research in Yunnan Province through grants 202301AS070047, 202301AT070471, and the Fundamental Research Funds for the Central Universities. 
Philip S. Yu was supported in part by NSF under grants III-1763325, III-1909323, III-2106758, and SaTC-1930941.

\bibliographystyle{ACM-Reference-Format}
\bibliography{sample-base}

\clearpage
\appendix
\section{Appendix}
\subsection{Details of Graph Learning Baselines}
\begin{itemize}[leftmargin=*]
    \item \textbf{LGCN}~\cite{lightgcn20}. This is the state-of-the-art recommendation method based on GCN~\cite{GCN16} by removing feature transformation and nonlinear activation.
    \item \textbf{DirectAU}~\cite{auloss}. This recommendation method proposes a novel loss based on alignment and uniformity of embedding distribution for model optimization.
    \item \textbf{UltraGCN}~\cite{ultragcn}. Instead of deploying graph convolution layers for recommendation, this method approximates infinite-layer graph convolutions via a constraint loss for message passing.
    \item \textbf{HGNN}~\cite{hnn}. It proposes hypergraph convolution for representation learning on hypergraphs. We calculate inner production between user and item nodes as the prediction scores with BPR loss as the objective for recommendation.
    \item \textbf{HCCF}~\cite{hccf}. In this hypergraph learning based recommendation method, contrastive learning is applied to jointly capture local and global collaborative relations.
    \item \textbf{DHCF}~\cite{dhcf}. It introduces hypergraph convolution into dual-channel learning for recommendation.
\end{itemize}
\subsection{Performance Curves in XMarket-CN and Steam datasets}
\begin{figure}[ht]

    \centering
    \includegraphics[width=0.43\textwidth]{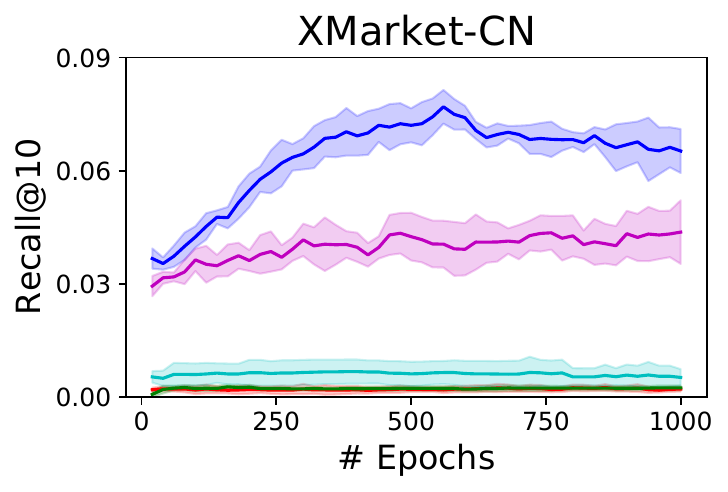}

    \centering
    \includegraphics[width=0.43\textwidth]{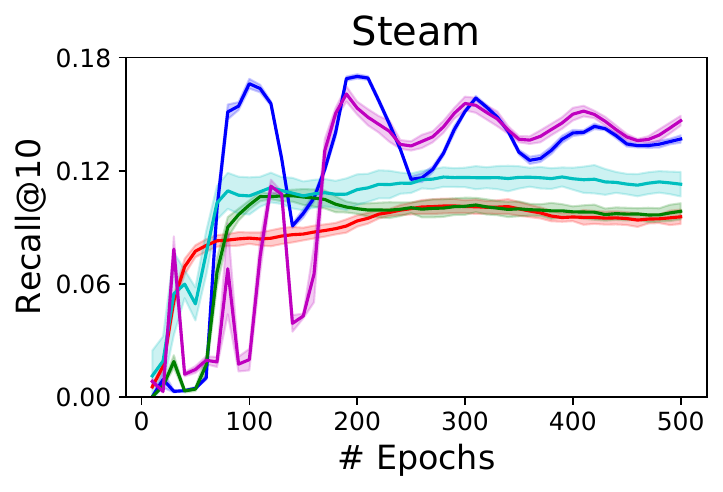}
    

  \caption{Testing recall w.r.t. the number of training epochs during finetuning in XMarket-CN and Steam datasets.}
\end{figure}

\end{document}